\documentclass[aip,apl,reprint,a4paper,floatfix,amsmath,amssymb,amsfonts,noshowpacs,citeautoscript]{revtex4-2}
\usepackage{newtxtext,newtxmath}
\usepackage[T1]{fontenc}
\usepackage{graphicx}
\usepackage[dvipsnames]{xcolor}
\usepackage{soul} 
\usepackage{floatrow}
\usepackage{xcolor} 

\usepackage[
,textwidth=17.5cm
,textheight=23.5cm
,verbose
,pdftex
]{geometry}
\usepackage{xspace}

\usepackage[pdftex]{hyperref}
\hypersetup{
  pdftitle={Dualtronics},
	colorlinks,
	citecolor=blue,
	linkcolor=blue
	}

\makeatletter
\newcommand*{\refig}[2]{\hyperref[#1]{\ref*{#1}(#2)}}
\makeatother

\usepackage{titlesec}

\DeclareMathAlphabet{\mathsf}{OT1}{\sfdefault}{m}{n}
\SetMathAlphabet{\mathsf}{bold}{OT1}{\sfdefault}{b}{n}

\usepackage{etoolbox} 
\makeatletter         

\newcounter{firstbib} 

\AtBeginEnvironment{thebibliography}{
}

\apptocmd{\thebibliography}{
  \setcounter{NAT@ctr}{\value{firstbib}}
}{}{}

\AtEndEnvironment{thebibliography}{
  \setcounter{firstbib}{\value{NAT@ctr}}
}

\begin{document}
\graphicspath{{./figures/}}

\title{Leveraging both faces of polar semiconductor wafers for functional devices}

\author{Len~van~Deurzen*}
\email[Author to whom correspondence should be addressed: ]{lhv9@cornell.edu}
\affiliation{School of Applied and Engineering Physics, Cornell University, Ithaca, New York 14853, USA}
\author{Eungkyun~Kim$^\dagger$}
\affiliation{Department of Electrical and Computer Engineering, Cornell University, Ithaca, New York 14853, USA}
\author{Naomi~Pieczulewski}
\affiliation{Department of Materials Science and Engineering, Cornell University, Ithaca, New York 14853, USA}
\author{Zexuan~Zhang}
\affiliation{Department of Electrical and Computer Engineering, Cornell University, Ithaca, New York 14853, USA}
\author{Anna Feduniewicz-Zmuda}
\author{Mikolaj~Chlipala}
\author{Marcin~Siekacz}
\affiliation{Institute of High Pressure Physics "Unipress", Polish Academy of Sciences, Warsaw 01-142, Poland}
\author{David~Muller}
\affiliation{School of Applied and Engineering Physics, Cornell University, Ithaca, New York 14853, USA}
\author{Huili~Grace~Xing}
\affiliation{Department of Electrical and Computer Engineering, Cornell University, Ithaca, New York 14853, USA}
\affiliation{Department of Materials Science and Engineering, Cornell University, Ithaca, New York 14853, USA}
\affiliation{Kavli Institute at Cornell for Nanoscale Science, Cornell University, Ithaca, New York 14853, USA}
\author{Debdeep~Jena}
\affiliation{School of Applied and Engineering Physics, Cornell University, Ithaca, New York 14853, USA}
\affiliation{Department of Electrical and Computer Engineering, Cornell University, Ithaca, New York 14853, USA}
\affiliation{Department of Materials Science and Engineering, Cornell University, Ithaca, New York 14853, USA}
\affiliation{Kavli Institute at Cornell for Nanoscale Science, Cornell University, Ithaca, New York 14853, USA}
\author{Henryk~Turski}
\affiliation{Department of Electrical and Computer Engineering, Cornell University, Ithaca, New York 14853, USA}
\affiliation{Institute of High Pressure Physics "Unipress", Polish Academy of Sciences, Warsaw 01-142, Poland}


\begin{abstract}
Unlike non-polar semiconductors such as silicon, the broken inversion symmetry of the wide bandgap semiconductor gallium nitride leads to a large electronic polarization along a unique crystal axis\cite{simon_polarization_doping}. This makes the two surfaces of the semiconductor wafer perpendicular to the polar axis dramatically different in their physical and chemical properties\cite{polarity_nanowires}. In the last three decades, the cation (gallium) face of gallium nitride has been used for photonic devices such as LEDs and lasers\cite{nobel_Nakamura,nobel_Akasaki,Nobel_Amano}. Though the cation face has also been predominantly used for electronic devices, the anion (nitrogen) face has recently shown promise for high electron mobility transistors due to favorable polarization discontinuities\cite{Npolar_HEMT_new}. In this work we introduce dualtronics, showing that it is possible to make photonic devices on the cation face, and electronic devices on the anion face, {\em of the same semiconductor wafer}. This opens the possibility for leveraging both faces of polar semiconductors in a single structure, where electronic, photonic, and acoustic properties can be implemented on opposite faces of the same wafer, dramatically enhancing the functional capabilities of this revolutionary semiconductor family.

\end{abstract}

\maketitle



Single crystals of semiconductors are macroscopic molecules with periodic repeating basis units\cite{brumfielElementalShiftKilo2010}. In a perfect semiconductor crystal wafer, the atomic location and type at one point in space repeats itself hundreds of microns from one surface to the other, and hundreds of millimeters from one edge to the other.  In non-polar semiconductors such as silicon, the crystal orientation is chosen carefully to maximize electronic performance of devices on the surface.  Electronic devices such as transistors are made only on one (say the top) surface of silicon and not the other.  Even if the other surface was used, the parallel surfaces of a cubic crystal are identical by symmetry.  This means flipping a silicon wafer upside down does not change the chemical or electronic properties of the top surface.  Thus, heterostructures and electronic devices made on the two surfaces in an identical manner exhibit identical properties.

\begin{figure*}[t]
\includegraphics[width=\textwidth]{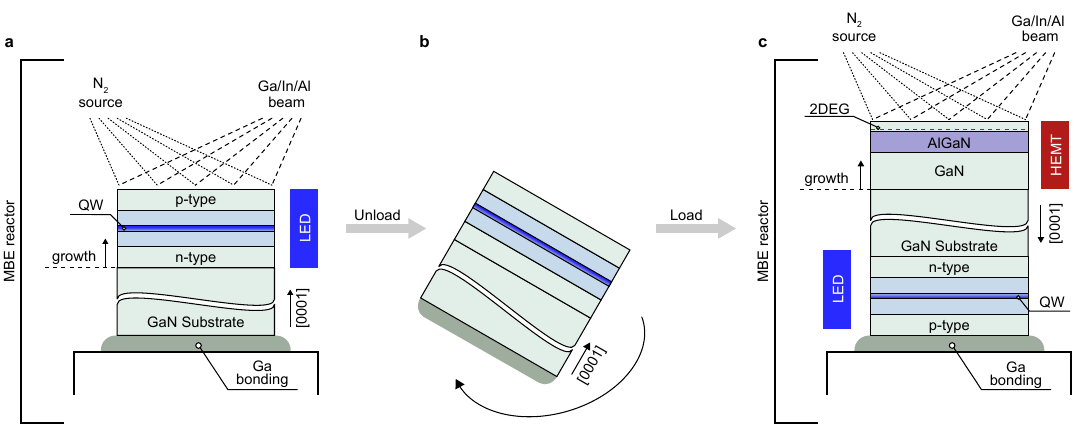}
\caption{Schematic of plasma-assisted molecular beam epitaxial growth of the HEMT-LED. The gray arrows indicate the chronological order of the growth procedure. (a) Ga-bonding of the GaN substrate and MBE growth of the (In,Ga,Al)N LED along the metal polar direction. (b) Unloading, cleaning, flipping, Ga-bonding and reloading of the sample. (c) MBE growth of the GaN/AlGaN HEMT along the nitrogen polar direction.  \label{Figure_1}}
\end{figure*}

The wide bandgap semiconductors Gallium Nitride (GaN) and Aluminum Nitride (AlN) have a wurtzite crystal structure with an underlying hexagonal lattice \cite{holmes-siedleGalliumNitrideValuable1974}. This crystal structure breaks inversion symmetry along the [0001] orientation (the c-axis).  The two surfaces perpendicular to the c-axis of a single crystal wafer of these polar semiconductors therefore exhibit dramatically different physical properties: flipping the crystal wafer is analogous to flipping a bar magnet. The chemical properties of the two sides is so distinct that they are used to identify metal or nitrogen surface polarity. While the metal-polar surface is inert to most chemicals, the N-polar surface of both GaN and AlN etches vigorously in solutions with bases, such as KOH and TMAH, or acids such as H${_3}$PO$_{4}$ \cite{WangAPL2013, jang2018polarity}. 

Dramatic differences in electronic properties of heterostructures are observed for the two faces \cite{Ambacher2DEGBothPolar,woodPolarizationEffectsSemiconductors2008}. For example, if a $\sim$10~nm thin coherently strained epitaxial GaN layer is deposited on the N-polar surface of AlN, a two-dimensional (2D) electron gas is formed at the GaN/AlN heterojunction quantum well due to the discontinuity of the conduction band combined with the discontinuity of the electronic polarization across the heterojunction \cite{zhangHighdensityPolarizationinduced2D2022}. But when an identical $\sim$10~nm coherently strained epitaxial GaN layer is deposited on the metal-polar surface of AlN, a 2D hole gas is formed at the heterojunction quantum well \cite{chaudhuriPolarizationinduced2DHole2019}. These polarization-induced, electrically conductive 2D electron and hole gases are formed in crystals nominally free of chemical impurities such as donor or acceptor dopants. (Al,Ga)N/GaN high electron mobility transistors (HEMTs) using such polarization-induced conductive channels demonstrate outstanding performance in high-power and high-speed applications \cite{zhengGalliumNitridebasedComplementary2021,40Wmm,GradedChannelWband,ShinoharaHighSpeed}.

To date, only a single face of c-axis oriented GaN single crystal wafers is used for either photonic, or electronic devices \cite{khanUltravioletLightemittingDiodes2008,pimputkarProspectsLEDLighting2009,emotoWidebandgapGaNbasedWattclass2022}. In this work, the two polarities on the opposite faces of a GaN single crystal are combined to realize a photonic device on one side, and an electronic device on the other. A heterostructure quantum well on the N-polar side is used to generate a polarization-induced high mobility 2D electron gas (2DEG), and a quantum well (In,Ga,Al)N pn diode heterostructure is realized on the opposite metal-polar side.  This two-sided wafer is then processed, first into HEMTs on the N-polar side, followed by blue quantum well light emitting diodes (LEDs) on the metal-polar side.  Successful operation of the HEMT and LED devices is observed, allowing the switching and modulation of blue LEDs by HEMTs on the other side of the wafer.

Figure \ref{Figure_1} shows how dual side epitaxy for dualtronics was achieved. The starting wafer is a high transparency bulk n-type O-doped GaN substrate grown by ammonothermal method \cite{grabianskaRecentProgressBasic2020,hashimotoGaNBulkCrystal2007} with mobile electron concentration $\sim$10$^{18}$~cm$^{-3}$, and threading dislocation density $\sim$10$^4$~cm$^{-2}$ with both the Ga-polar and N-polar sides chemo-mechanically polished to obtain atomically smooth surfaces. The N-polar surface of the GaN substrate was gallium-bonded to a GaN/sapphire carrier wafer as shown in Figure~\ref{Figure_1}{(a)}.  Because the surface bonded to the carrier wafer needs to be used again, we chose gallium bonding to the GaN surface of a GaN/sapphire carrier wafer to ensure desired elements (Ga,~N) at this bonding interface.  Using molecular beam epitaxy (MBE), we first grew a blue (In,Ga,Al)N LED structure consisting of a quantum well active region inside a pn heterojunction on the Ga-polar face of GaN.  The wafer was then detached from the carrier wafer and flipped as shown in Figure \ref{Figure_1}{(b)}. The Ga was removed from the N-polar side, and the metal-polar p-type GaN surface of the LED side was gallium-bonded on to a clean GaN/sapphire carrier wafer. Subsequently, a N-polar GaN/AlGaN/GaN heterostructure was grown on the N-polar face again by MBE to create a high-electron mobility 2DEG to serve as the conduction channel of the HEMT as shown in Figure \ref{Figure_1}{(c)}.  The chosen growth conditions for the LED and HEMT structures are described in the methods section.

\begin{figure*}[t]
\includegraphics[width=\textwidth]{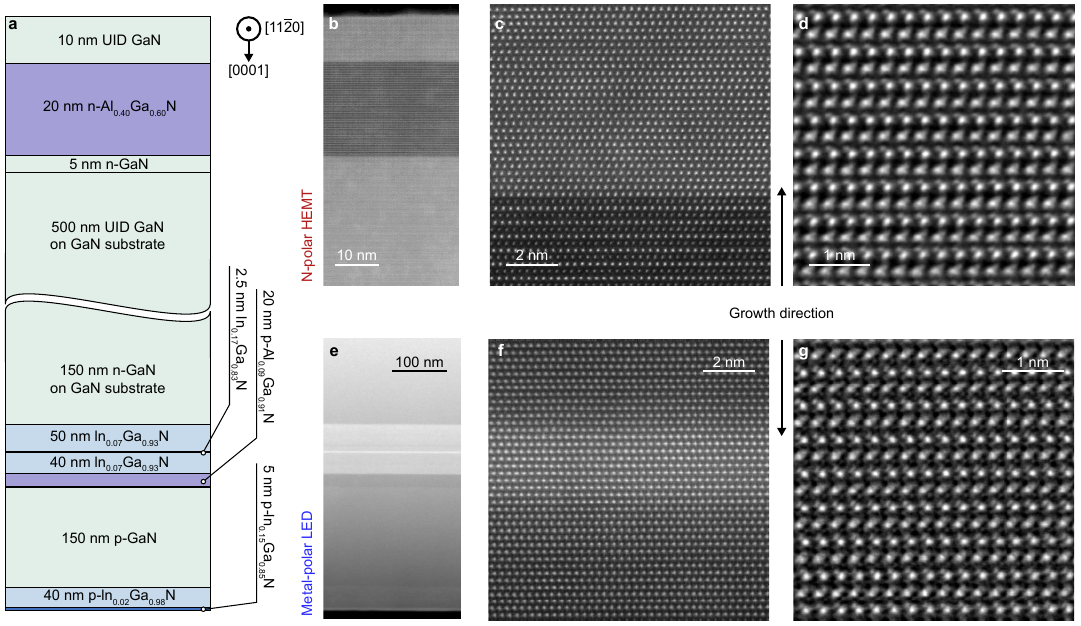}
\caption{STEM imaging of dualtronic epitaxial heterostructure. (a) Schematic of the HEMT-LED heterostructures grown on both faces of a single crystal c-plane n-GaN substrate. (b) HAADF-STEM image showing the GaN/Al$_{0.40}$Ga$_{0.60}$N/GaN HEMT and (c) atomic resolution image corresponding to the uppermost GaN/Al$_{0.40}$Ga$_{0.60}$N heterojunction interface which hosts the 2DEG. (d) iDPC image in the uppermost GaN layer of the HEMT indicating that the nitrogen polarity follows that of the substrate to the surface. (e) HAADF-STEM image showing the LED quantum well, electron blocking layer, cladding and contact layers and (f) atomic resolution image corresponding to the LED In$_{0.07}$Ga$_{0.93}$N/In$_{0.17}$Ga$_{0.83}$N/In$_{0.07}$Ga$_{0.93}$N single quantum well. (g) iDPC image of the p-InGaN contact layers of the LED indicating the metal polarity follows that of the substrate.
\label{Figure_2}}
\end{figure*}

Figure \ref{Figure_2}{(a)} shows the precise layer thicknesses and compositions of the HEMT on the N-polar side on the top and the LED on the metal-polar side on the bottom of the single GaN wafer achieved after the double side epitaxy.  Figures~\ref{Figure_2}{(b-g)} show the corresponding scanning transmission electron microscopy (STEM) images. 

The MBE grown HEMT on the N-polar side started with the growth of a 500 nm thick unintentionally doped (UID) GaN buffer layer as shown in Figure \ref{Figure_2}{(a)}, followed by 5~nm GaN:Si and 5 nm Al$_{0.40}$Ga$_{0.60}$N:Si layers with a Si concentration of 3~$\times$~$10^{18} \mathrm{cm}^{-3}$ capped with undoped 15~nm Al$_{0.40}$Ga$_{0.60}$N and 10~nm GaN. Silicon doping below the 15~nm AlGaN layer was employed to prevent the formation of a polarization-induced 2D hole gas at the bottom N-polar AlGaN/GaN interface\cite{RajanNpolar}.  The bulk energy bandgap of GaN is $E_{g1}=3.4$ eV, and of Al$_{0.40}$Ga$_{0.60}$N is {$E_{g2}=4.5$ eV}. For the N-polar HEMT the high-angle annular dark-field (HAADF) STEM image in Figure \ref{Figure_2}{(b)} shows the 20~nm AlGaN layer near the surface with sharp interfaces with GaN on both sides. Figure \ref{Figure_2}{(c)} zooms in to show the lattice image of the top GaN/AlGaN heterojunction where the 2DEG resides on mostly the GaN side. A sharp heterointerface between GaN and AlGaN is observed with the atomic planes clearly resolved.  Figure \ref{Figure_2}{(d)} shows an integrated differential phase contrast (iDPC) image of the GaN HEMT channel layer, proving the N-polar crystal structure along the growth direction: the N-atoms located vertically below the brighter Ga atoms.

For a N-polar heterostructure, a mobile 2DEG is expected to form on the GaN side inside a triangular quantum well formed by the conduction band offset {$\Delta E_{c} \approx 0.8$}~eV at the sharp GaN/AlGaN heterojunction seen in Figure \ref{Figure_2}{(c}). A Hall-effect transport measurement described in the methods section revealed that a 2DEG is formed with room-temperature electron density $n_s \approx $ 1.26 $\times$ 10\textsuperscript{13} cm$^{-2}$, and electron mobility of $\mu \approx $ 1970~cm\textsuperscript{2}/(V$\cdot$s), resulting in a sheet resistance of $R_{sh} = 1/(e n_{s} \mu) \approx $ 252 $\Omega$/sq, where $e$ is the magnitude of the electron charge. The observed sheet resistance is the lowest and the electron mobility is one of the highest among all N-polar HEMT heterostructures reported in the literature, resulting largely from low interface roughness scattering of the quantum confined electrons at the sharp interface due to the high quality epitaxy on the double-side polished bulk GaN substrate \cite{DiezNpolarRecord,PasayatFirstNpolarOnBulkGaN,BrownNpolarGrowth}. A benchmark comparison with state-of-the-art N-polar III-nitride HEMTs reported in the literature is shown in Extended Data~Fig.~1.

The MBE grown LED heterostructure on the metal-polar side seen on the bottom side of Figure \ref{Figure_2}{(a)} consists of a 150~nm GaN:Si buffer layer followed by a 50~nm In$_{0.07}$Ga$_{0.93}$N/40~nm In$_{0.07}$Ga$_{0.93}$N active region with a single 2.5~nm thick In$_{0.17}$Ga$_{0.83}$N quantum well embedded inbetween.

This photonic active region was followed by p-type doped layers starting with a 20 nm Al$_{0.09}$Ga$_{0.91}$N electron blocking layer with a Mg concentration of 3~$\times$~$10^{19} \mathrm{cm}^{-3}$ followed by a 150~nm thick GaN layer with a Mg concentration of 6~$\times$~$10^{18} \mathrm{cm}^{-3}$.  This LED heterostructure was capped with p-InGaN contact layers consisting of 40~nm thick In$_{0.02}$Ga$_{0.98}$N and 5~nm thick In$_{0.15}$Ga$_{0.85}$N doped with Mg at the level of 1~$\times$~$10^{20} \mathrm{cm}^{-3}$ and 5~$\times$~$10^{20} \mathrm{cm}^{-3}$, respectively, to form low resistance ohmic contacts for hole injection into the LED active region, while simultaneously to be transparent to the light emitted from the active region.  Figure \ref{Figure_2}{(e)} shows a HAADF-STEM image of each of these LED layers indicating a uniform quantum well and the surrounding active region along the plane of the wafer.  The atomically resolved image in Figure~\ref{Figure_2}{(f)} shows the 2.5~nm thick In$_{0.17}$Ga$_{0.83}$N quantum well and the InGaN cladding layers. The group III element fractions and doping profiles for both the LED and HEMT structure are presented in the time-of-flight secondary ion mass spectrometry (ToF-SIMS) data shown in Extended Data Fig.~2, which indicate that the measured values are as intended. Importantly, there is no noticeable interdiffusion of dopants or group-III broadening of the metal-polar LED layers, indicating interface and doping control is not affected by the second growth step.  

\begin{figure*}[t]
\includegraphics[width=\textwidth]{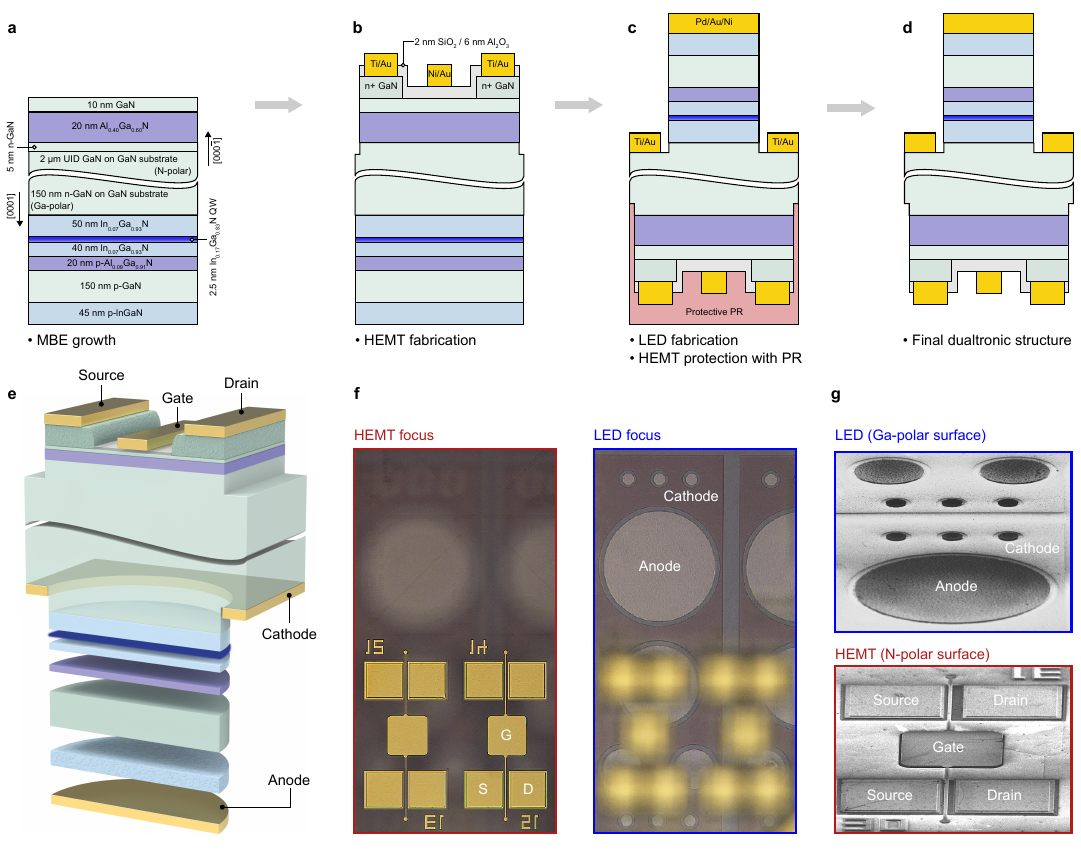}
\caption{Fabrication and imaging of dualtronic device. (a)-(d) Device processing flow for the double-sided HEMT-LED. Starting from the as-grown heterostructures, the gray arrows follow the independent processing steps chronologically, with the metal-polar LED being processed after the N-polar HEMT. (e) A three-dimensional representation of the complete device. (f) Optical microscope images of the as-processed sample, focused on the LEDs (right) and the HEMTs (left). The N-polar HEMTs are oriented upwards, forming the uppermost surface. For scale, the diameter of the large LED anode contact is 140~\textmu m. (g) Scanning electron microscope (SEM) images of the HEMTs on the N-polar GaN surface (bottom) and the LEDs on the metal-polar GaN surface (top). \label{Figure_3}}
\end{figure*}

Figure \ref{Figure_2}{(g)} shows an iDPC image of the p-InGaN contact layers on the metal-polar LED side. The LED is indeed metal-polar viewed in the growth direction with the substrate out-of-view above. This confirms that the polarity of the crystal substrate locks the polarity of the epitaxial layers, and that the crystalline registry goes through the entirety of the GaN wafer over hundreds of microns.  This is why the 2DEG forms on the the N-polar side of the wafer, while the metal-polarity is retained throughout the active regions, quantum well, and electron blocking and contact layers of the LED.  The high crystalline perfection of the LED heterostructure also proves that the growth of the HEMT heterostructure on the N-polar side at a high temperature and with nitrogen plasma does not destroy or degrade the LED structure during the rather harsh thermal and chemical conditions. The surface of the as-grown heterostructures after the completion of both epitaxial steps as shown in Extended Data Fig.~3 exhibit sub-nanometer roughness atomic steps characteristic of step-flow growth, indicating the successful dual side epitaxy. X-ray diffraction measurements indicate high crystallinity, sharp heterointerfaces, and coherently strained layers to the GaN substrate seen in Extended Data Fig.~4.

\begin{figure*}[t]
\includegraphics[width=\textwidth]{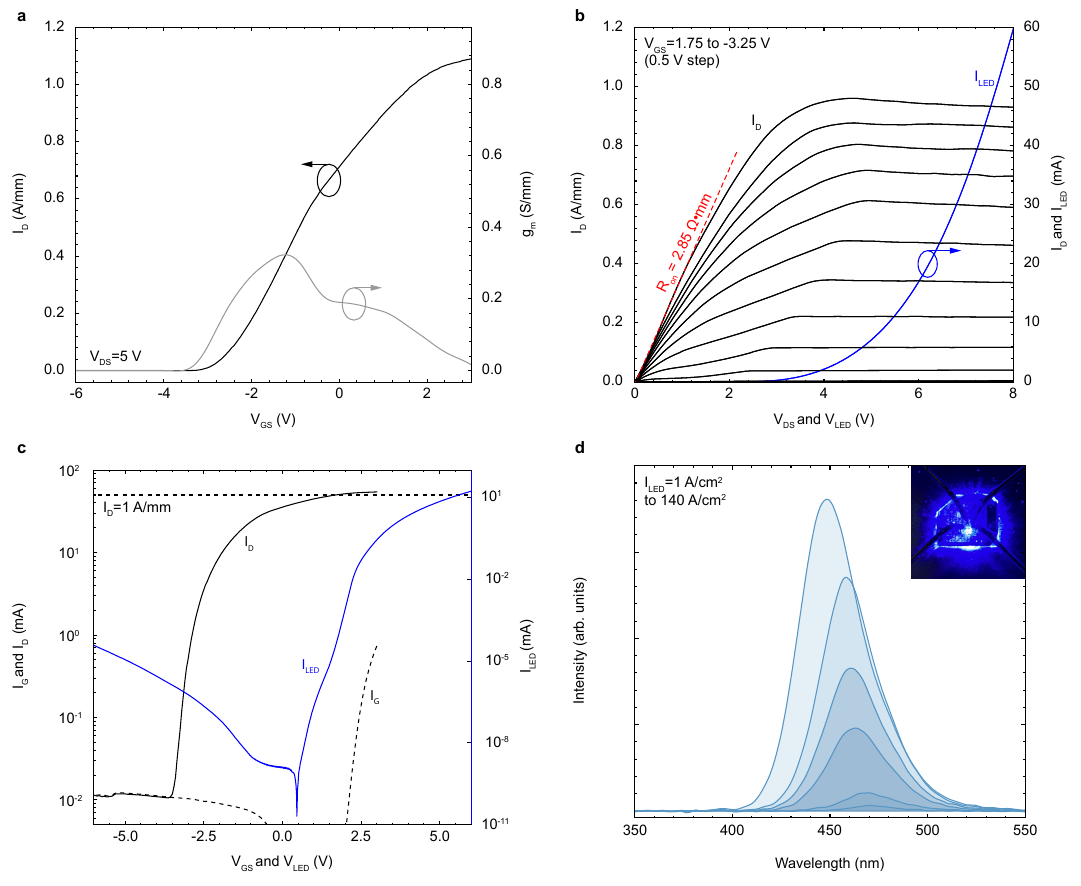}
\caption{Device characteristics of the HEMTs and LEDs, operating independently. (a) Normalized drain current (black, solid) and transconductance (gray, solid) of an N-polar HEMT as a function of gate-source voltage, operating at a drain-source voltage of 5~V. (b) Linear plot showing the family of curves for a HEMT (solid black) for a gate-source voltage ranging from 1.75 (on) to -3.25~V (off). On the right axis, the linear current-voltage characteristics of a 400~\textmu m diameter LED (blue, solid) and unnormalized output characteristics of a HEMT are shown as well. The dimensions of the measured HEMTs are L\textsubscript{SD} = 4 \textmu m, L\textsubscript{G} = 1.5 \textmu m, and W\textsubscript{G} = 50 \textmu m. (c) Semi-log plots showing the unnormalized drain current (black, solid) and gate current (black, dashed) versus gate-source voltage, for a drain-source voltage of 5~V. Here the transistor current corresponds to the left vertical axis. The horizontal, black, dashed line indicates a normalized channel sheet current-density of 1 A/mm. Similarly, corresponding to the right vertical axis is the LED current (blue, solid) as a function of forward bias for a 400~\textmu m diameter device. (d) Electroluminescence spectra of a metal-polar, 400~\textmu m diameter LED. The injection current density ranges from 1~to~140~A/cm$^{2}$. The top-right inset shows a camera image of the sample with an LED in the on-state.
\label{Figure_4}}
\end{figure*}

For successful dualtronics, the survival of the thin LED heterostructure layers during the epitaxial growth conditions of the HEMT side needs to be replicated for the steps of device fabrication.  The N-polar surface is particularly difficult to protect chemically because it reacts strongly with several solutions that are typically used for surface cleaning and conditioning during the processing steps.  For example, the natural etching of the N-polar surface of GaN in solutions of KOH, TMAH, or H$_{3}$PO$_{4}$ form pyramidal facets that significantly roughen the pristine, flat surface \cite{wang2013n}.  This property of surface roughening is put to good use for efficient light extraction in blue LEDs.  But such roughening would destroy the N-polar HEMT desired here, which requires sub-nanometer smoothness to preserve the high mobility 2DEG and for the fabrication of source, drain, and gate electrodes.  Figure \ref{Figure_3} shows the procedure adopted for the fabrication of the HEMT and the LEDs on the opposite sides of the same wafer, along with the final dualtronic device. A more detailed overview schematic of the fabrication of the dualtronic device is shown in Extended Data Fig.~5.

\begin{figure*}[t]
\includegraphics[width=\textwidth]{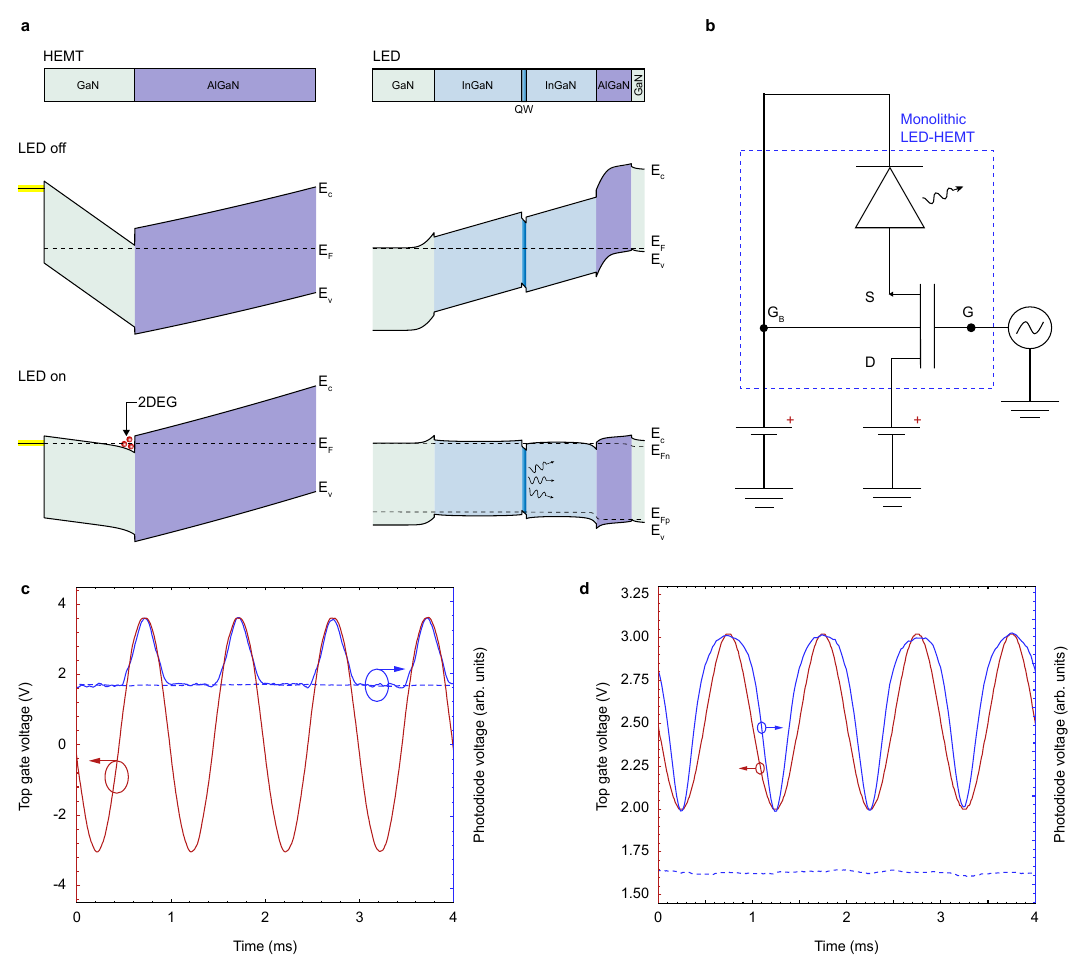}
\caption{Monolithic HEMT-LED switching measurements. (a) Energy-band diagrams of the HEMT and LED indicating the off and on-states. (b) Circuit schematic of the monolithic HEMT-LED, taking into account the back-gating effect of the conductive GaN substrate. (c)-(d) Monolithic switching measurements, modulating between on-off (c) and between bright and dim modes (d). The gate-source modulation voltage is shown in red, the photodiode voltage while modulating the LED in solid blue, and the background (LED off state) photodiode voltage in dashed blue.  \label{Figure_5}}
\end{figure*}

Figure \ref{Figure_3}{(a)} shows the starting layer structure which is first processed into a HEMT on the N-polar side shown in Figure \ref{Figure_3}{(b)}.  Ti/Au low resistance source and drain ohmic contacts are formed on regrown GaN layers that are heavily n-type doped with Silicon.  A thin gate SiO$_2$/Al$_2$O$_3$ dielectric layer was deposited followed by a Ni/Au gate metal stack to complete the HEMT fabrication.  The HEMT layer was then protected by a photoresist coating and flipped upside down to the metal-polar side up for LED fabrication for each photolithography step.

Figure \ref{Figure_3}{(c)} shows that for the fabricated LED, Pd/Au metallization was used for the p-type anode contact. The high work function metal Palladium aligns the metal Fermi level close to the valence band holes of the p-type doped InGaN cap layer of the LED side.  The thin p-InGaN reduces the contact resistance by enhancing the free hole concentration and by lifting the valence band energy-maximum.  To form the n-type cathode contact, mesa regions were etched by removing the entire active region and terminating the etch on the buried n-type GaN layer to reveal a flat surface.  Ti/Au metal stacks were deposited on this flat n-GaN surface to form a low resistance contact between electrons at the Fermi surface of the metal and the electrons in the GaN conduction band.   

Figure~\ref{Figure_3}{(d)} shows the completed dualtronic structure after removal of the protective photoresist coating, and Figure \ref{Figure_3}{(e)} shows a cross-section of it flipped upside down.  This schematic emphasizes the single-crystal nature of the whole structure, across the bulk of the wafer indicated by the wavy break in the thick GaN layer between the HEMT on the top and the LED on the bottom.  Figure~\ref{Figure_3}{(f)} shows an optical microscope image focused on the top HEMT surface on the left, and the LED focused on the right.  The energy bandgap of the bulk wafer between the HEMT and LED mesas is $E_{g1}=3.4$~eV of GaN, which is optically transparent to visible light, allowing for imaging of the LEDs through the wafer from the HEMT side as seen in Figure~\ref{Figure_3}{(f)}.  This also emphasizes that the transistor layers are transparent to visible wavelengths, and are therefore attractive for dualtronic integration with LEDs as transparent thin-film transistors (TFTs) but, as seen later, with dramatically higher performance than existing TFTs made of oxide or organic semiconductors. Figure \ref{Figure_3}{(g)} shows the scanning electron microscope (SEM) images of the HEMTs on the N-polar GaN surface (bottom) and the LEDs on the metal-polar GaN surface (top).  

Figure \ref{Figure_4}{(a)} shows the measured transfer characteristic and transconductance of the N-polar HEMT. A threshold voltage of $V_{T} \approx -3$~V is measured with an on-current drive exceeding 1~A/mm at a drain voltage of $V_{DS}=5$~V and gate voltage $V_{GS} \approx 3$~V. The transistor transconductance $g_{m} = \partial I_{D}/\partial V_{GS}$ peaks at $V_{GS} \approx -1.6$~V. Figure \ref{Figure_4}{(b)} shows the measured $I_{D} - V_{DS}$ output characteristics of the HEMT for various gate voltages overlaid with the LED current-voltage characteristics.  The transistor delivers enough output current to drive the LED with variable drive currents controlled by the gate voltage up to 50~mA or more. Figure \ref{Figure_4}{(c)} shows the LED current and the HEMT current for $V_{DS}=5$~V in the logarithmic scale indicating that the HEMT can control the LED current and hence its optical output intensity over several orders of magnitude. Figure \ref{Figure_4}{(d)} shows the measured output spectra of the LED for injection current densities ranging from 1 to 140~A/cm$^2$.  The emission intensity increases with the injection current density, and the peak emission wavelength blue-shifts from $\lambda \approx 470$~nm at 1 A/cm$^2$ to $\lambda \approx 450$~nm at 140~A/cm$^2$ in the blue regime of the visible spectrum. The blueshift is caused by screening of the internal polarization fields which result in a reduced quantum confined Stark effect (QCSE), as well as the Burstein-Moss effect. The inset in Figure \ref{Figure_4}{(d)} shows the bright blue electroluminescent glow of the blue LED slightly below the center of the wafer on the metal-polar side.  The output and transfer characteristics of the HEMTs before and after the fabrication of the LEDs is shown in the Extended Data Fig.~6. It was confirmed that after the LED processing, there was minimal degradation of the HEMT with a less then 0.3~V threshold voltage shift and negligible change in the output and gate leakage current. This minimal degradation highlights the feasibility of using double-side epitaxy and processing for reliable heterogeneous device integration on both crystal faces.

Figure \ref{Figure_5}{(a)} shows the layer structures and corresponding energy band diagrams of the HEMT on the left and the LED on the right when they are connected in the circuit format shown in Figure \ref{Figure_5}{(b)}.  The source of the HEMT is connected to the anode of the LED to inject current to light up the LED.  The light emitted from this dualtronic device is measured with a photodiode. A customized probe setup was fabricated to probe both HEMTs and LEDs without the need to flip the sample, shown schematically in Extended Data Fig~7. When the gate voltage of the HEMT is below the transistor threshold ($V_{GS}< V_T$), the triangular quantum well is lifted above the Fermi level in the energy band diagram of the HEMT as shown, and the transistor is off.  The corresponding energy band diagram of the LED is shown on the right.  When the HEMT gate voltage is above threshold, the bottom of the triangular quantum well is pulled below the Fermi level as seen in Figure \ref{Figure_5}{(a)}, flooding the GaN/AlGaN interface with electrons that form a 2DEG channel. The output current flowing from the drain of the HEMT is then injected into the cathode of the LED, forward biasing it to the energy band diagram shown and indicated by the split quasi-Fermi levels of electrons $E_{fn}$ and holes $E_{fp}$.  The electrons injected from the cathode radiatively recombine with the holes injected from the anode into the In$_{0.17}$Ga$_{0.83}$N quantum well, emitting light that is sensed by the photodiode. 

Figure \ref{Figure_5}{(c)} shows that the photovoltage signal follows the gate voltage as it is swept between $ -3.0 \leq V_{\mathrm{GS}} \leq$~3.6~V with the drain voltage set at a voltage of 5.9~V and the backgate voltage set at a voltage of 1~V, over millisecond timescale across its threshold voltage from on state to off state. Figure \ref{Figure_5}{(d)} shows that when the gate voltage of the HEMT is modulated between $2.0 \leq V_{\mathrm{GS}} \leq 3.0$~V when it is in its on state with $V_{\mathrm{GS}}>V_{\mathrm{T}}$, the light emitted by the LED follows this gate modulation, translating the electronic signal in the gate of the HEMT into a photonic bright and dim signal output from the LED on the other side of the wafer.  The on-off modulation of the dualtronic device can be improved by utilizing similar concepts as used for traditional LEDs. \cite{Lu_bandwidth,Chai_bandwidth}

Due to the dualtronic structure, the cathode of the LED can also serve as a back-gate for the HEMT, which was taken into account in the monolithic switching measurements. With a separate contact to the n-GaN substrate we observed that the cathode voltage could exponentially control the drain current when the top gate was electrically floating. This back gating effect is available as a new functionality, or may be eliminated if undesirable by replacing the conductive substrate with a semi-insulating substrate. The backgating effect of the HEMT with the LED is shown in Extended Data Fig.~8. 

The observations described here thus prove that the concept of dualtronics is feasible, opening the gates for many interesting possibilities.  This work realized an electronics fabric on the N-polar face and a photonic fabric on the metal-polar face of the same wafer.  This dualtronic combination is of immediate interest for combining microLEDs on the photonic side with transparent TFTs on the same GaN wafer on the other side. This monolithic convergence of devices reduces the number of components required for microLEDs with the potential for dramatic area and cost savings due to efficient use of the substrate real-estate. Moreover, monolithic integrating schemes combining transistors and LEDs on a single substrate face rely on multiple epitaxial layers that need to be either selectively removed and/or regrown in order to expose the LED heterostructure buried underneath the transistor structure or vise versa \cite{liuMonolithicIntegrationAlGaN2014,KalaitzakisMonolithicIntegration,LiMonolithicIntegration,LiMonolithicIntegration2,ChenMonolithicIntegration,nano24,rahman2024,bharadwajBottomTunnelJunction2020}. Exposing the buried heterostructure by dry etching induces plasma damage, leading to diminished light emission and degradation of contacts. Similarly, with a selective growth method, the ex-situ nature of regrowth results in a poor growth interface, increasing leakage pathways. These issues are all avoided with the dualtronics integration scheme.

But the dualtronics concept extends to several exciting new opportunities. The metal-polar face of the substrate, to take advantage of high emission efficiencies, can be utilized for any optoelectronic device like laser diodes, semiconductor optical amplifiers (SOAs), and electro-optical modulators, while transistors or photodetectors are fabricated on the N-polar face. This full utilization of the substrate dramatically decreases the number of components and chips needed in photonic integrated circuits. For other applications, both GaN polarities can be used. For instance, RF transistor power amplifiers (PAs) for the transmit part of communication systems can be realized on one polarity, and low-noise amplifiers (LNAs) for the receive end of communication systems can enable integrated transceivers in a combined transmit/receive module of smaller form factor than existing systems. The combination of n-channel transistors on one polarity with p-channel transistors on the opposite polarity can enable new forms of complementary transistor circuit topologies connected by through-vias through the substrate. Such dualtronic devices can take advantage of the wide bandgap nature of the polar nitride semiconductors for new forms of power electronics and RF electronics. These, and several allied possibilities, can allow for the creation and manipulation of electrons and photons on the opposite faces of the same wafer to achieve new functionalities.

The ultra wide bandgap polar semiconductor AlN boasts large electro-acoustic coupling, which makes it the preferred material today for acoustic wave RF filters.  Dualtronics can thus take advantage of this property of the polar nitride semiconductors to combine sonar (by sound waves), radar (microwaves), and lidar (light) on the same platform.

The efficient use of the substrate surfaces eliminates wasted space, reduces the energy and material costs of producing multiple wafers, and thus should be of high interest for future technologies well beyond the particular polar semiconductor materials discussed here.

\bibliography{ms}

\pagebreak

\clearpage


\titleformat*{\section}{\Large\bfseries}

\section*{Methods}

\titleformat*{\section}{\normalfont\bfseries}

\section*{Preparation of bulk GaN substrate}
The investigated sample was grown on a bulk O-doped GaN 1~$\times$~1 $\mathrm{cm}^{2}$ high transparency substrate grown by an ammonothermal method with a room-temperature free electron concentration reaching 10$^{18}$~cm$^{-3}$ and threading dislocation density on the order of 10$^4$~cm$^{-2}$ \cite{zajac2018}. Using the high transparency substrate allows for better light extraction and device characterization. Both faces of the substrate were first mechanically polished to obtain flat and parallel surfaces, both profiting from a typically used miscut angle of 0.5$^{\circ}$ of the (0001) plane towards [1$\bar{1}$00]. Both the Ga-polar and N-polar surfaces were then chemo-mechanically polished to obtain atomically smooth surfaces exhibiting atomic steps.

\section*{Molecular beam epitaxy of metal-polar LED and nitrogen-polar HEMT}
The epitaxial processes on both crystal planes were conducted using plasma-assisted molecular beam epitaxy in a Veeco Gen20a reactor. GaN and AlGaN layers on both polarities were grown at the same substrate temperature around 730$^{\circ}$C using a growth rate of 0.4~\textmu m/h (realized by nitrogen plasma source operating at 250W and 2.2 sccm N$_{2}$ flow), while InGaN layers were grown at a temperature of 650$^{\circ}$C using a growth rate of 0.8~\textmu m/h (plasma source operating at 310W and 10 sccm N$_{2}$ flow). The growth temperature for GaN and InGaN was deduced using reference fluxes of Ga and In desorption rates, respectively \cite{siekacz2009ingan}. Desorption rates were calibrated as a function of the real substrate temperature in a separate process on a 2-inch GaN/Al$_2$O$_3$ reference wafer using a k-Space BandiT pyrometer that measures temperature using the principle of blackbody radiation. The Ga-polar and N-polar structures were both grown using commonly used metal-rich growth conditions similar to recipes described elsewhere for laser diode growth \cite{skierbiszewski2014nitride}. 
The sequence of the epitaxial procedures, as well as the growth conditions and duration, can potentially affect the performance of the devices. In this study, the Ga-polar LED was grown first to avoid the risk of degrading the N-polar HEMT device performance.

The LED heterostructure was grown starting with a 150~nm GaN:Si buffer layer followed by an In$_{0.17}$Ga$_{0.83}$N/In$_{0.07}$Ga$_{0.93}$N active region with a single 2.5~nm thick In$_{0.17}$Ga$_{0.83}$N quantum well. The active region was followed by p-type doped layers starting with an Al$_{0.09}$Ga$_{0.91}$N electron blocking layer with a Mg concentration of 1~$\times$~$10^{19} \mathrm{cm}^{-3}$ followed by a 150~nm thick GaN layer with a Mg concentration of 4~$\times$~$10^{18} \mathrm{cm}^{-3}$,  finally capped with p-InGaN contact layers consisting of 40~nm thick In$_{0.03}$Ga$_{0.97}$N and 5~nm thick In$_{0.15}$Ga$_{0.85}$N doped with Mg at the level of 4~$\times$~$10^{19} \mathrm{cm}^{-3}$ and 5~$\times$~$10^{20} \mathrm{cm}^{-3}$, respectively. After the LED growth, the sample was taken out of the reactor and etched using a piranha solution at 60$^{\circ}$~C to remove residual gallium without affecting the epi-ready state of the N-polar surface. To confirm that the surface morphology of the LED is not affected by the secondary N-polar growth procedures, including gallium mounting, HEMT growth, and wet chemical piranha cleaning, AFM was performed on the $+c$ LED surface of the sample before and after these secondary growth procedures. The 2~$\times$~2 \textmu m$^{2}$ AFM topographs in Extended Data Figs.~9(a)~and~(b) corroborate these findings.

Subsequently, the N-polar HEMT heterostructure was grown, which started with a 500~nm thick undoped GaN buffer layer followed by 5~nm GaN:Si and 5~nm Al$_{0.40}$Ga$_{0.60}$N:Si layers with a Si concentration of 3~$\times$~$10^{18} \mathrm{cm}^{-3}$ capped with undoped 15~nm Al$_{0.40}$Ga$_{0.60}$N and 10~nm GaN. Silicon doping in the preceding layers was used to prevent the formation of a two-dimensional hole gas at the bottom interface. The two-dimensional electron gas (2DEG) in such structure forms in the top-most unintentionally doped GaN layer. After N-polar growth, the same cleaning recipe involving a pirhana etch was used to remove residual galllium from the Ga-polar surface while not affecting the N-polar surface.

\section*{Structural characterization of the LED and HEMT heterostructures}
Depth-resolved concentrations of Mg, O, C, and Si,  as well as the group III element fractions were measured by time-of-flight secondary ion mass spectrometry (ToF-SIMS) performed by Evans Analytical Group. Measurements were performed on both faces separately into the substrate and are shown in Extended Data Fig.~2.

Scanning transmission electron microscopy (STEM) was performed on the samples, which were prepared using a Thermo Fisher Helios G4 UX Focused Ion Beam with a final milling step of 5~keV. STEM measurements were taken on both faces independently with an aberration-corrected Thermo Fisher Spectra 300 CFEG operated at 300~keV. Corresponding high-angle annular dark-field (HAADF) images and integrated differential phase contrast (iDPC) images of the HEMT and LED heterostructures are shown in Fig.~2.

Atomic force microscopy (AFM) was performed with an Asylum Cypher AFM microscope in tapping mode for sample areas of 5~×~5~\textmu m$^2$ and 2~×~2~\textmu m$^2$. Atomic force topographs of the $+\mathrm{c}$ and $-\mathrm{c}$ surfaces of the as-grown heterostructures are shown in the Extended Data Fig.~3.

Triple-axis $\omega-2\theta$ x-ray diffraction (XRD) scans across the symmetrical 002 wurtzite reflections were measured using a Panalytical Empyrean x-ray diffractometer equipped with a PIXcel$^{3\mathrm{D}}$ detector and Xe proportional detector. The monochromator consists of a hybrid two-bounce Ge220 crystal utilizing Cu~${\mathrm{K}\alpha_{1}}$ radiation. Due to the thick ~400 \textmu m substrate, two independent scans were performed for the nitrogen-polar HEMT face and metal-polar LED face, as shown in Extended Data Fig.~4.

\section*{Fabrication of nitrogen-polar HEMTs}
Device fabrication commenced with processing of the HEMT on the N-polar side of the double-sided sample, since this face is much more reactive than the metal-polar face. First, source and drain ohmic contacts were defined by patterning the N-polar heterostructure with a 200/30~nm plasma enhanced chemical vapor deposition SiO\textsubscript{2}/electron-beam evaporated Cr hard mask. After removing the Cr and SiO\textsubscript{2} layers in the contact regions via a selective dry etch to expose the GaN channel layer, the sample was reloaded into the MBE chamber and 50~nm n\textsuperscript{++} GaN (N\textsubscript{D} $\sim$ 10\textsuperscript{20}cm\textsuperscript{-3}) was regrown at a thermocouple temperature of 660~$^{\circ}$C. The excess n\textsuperscript{++} GaN was lifted off by removing the SiO\textsubscript{2}/Cr hard mask in diluted HF, and a 6/2~nm Al\textsubscript{2}O\textsubscript{3}/SiO\textsubscript{2} gate dielectric was blanket deposited by thermal ALD, followed by post deposition annealing at 400~$^{\circ}$C in O\textsubscript{2}, as shown in Extended Data Fig.~5(b). Next, mesa isolation was achieved by a BCl\textsubscript{3} inductively coupled plasma reactive ion etch, which extended into the GaN buffer layer. Source and drain non-alloyed ohmic metallization regions were then defined by photolithography, and prior to the e-beam evaporation of 50/100 nm Ti/Au, a gate dielectric and a few nanometers of regrown n\textsuperscript{++} GaN layer were removed by a diluted HF dip and lower power ICP etch, respectively, to expose the clean surface, as shown in Extended Data Fig.~5(c). Lastly, rectangular gates shown in Extended Data Fig.~5(d) were defined by photolithography and metallized by e-beam evaporation of 50/100~nm Ti/Au. 

\section*{Fabrication of metal-polar LEDs}
After completion of the N-polar HEMTs, the metal-polar face was processed into light-emitting diodes (LEDs) in the manner as described in the bottom row of the extended Data Fig.~5(e)-(g). First, before each photolithography step, a positive photoresist was spun and baked on the nitrogen-polar face to protect and preserve the features of the HEMTs. Then, a Pd/Au/Ni metallization stack with thicknesses of 20/100/50~nm was deposited by electron beam evaporation, forming the anode (p-side electrode) for the LEDs. This stack was subsequently used for self-aligned etching and formation of the device mesas by BCl$_{3}$ inductively coupled reactive ion etching , resulting in an etch depth of approximately 540~nm. The device mesas are circular with a diameter ranging between 20 and 400~\textmu m. Finally, a Ti/Au cathode (n-side electrode) metallization stack was deposited at thicknesses of 20/100~nm by electron beam evaporation, completing the LED formation.

\section*{LED and HEMT device characterization}
On-wafer van der Pauw patterns were used for Hall-effect measurements and to extract the sheet carrier density and electron mobility of the 2DEG. Current-voltage curves obtained from linear transfer-length method patterns were used to extract the contact resistance of the HEMT source and drain contacts (from the metal pad to 2DEG) and sheet resistance, which matched well with the values obtained by Hall-effect measurements. Current-voltage characteristics of the LEDs and transfer and output curves for the HEMTs, operating independently from each other, were measured by using a Cascade Microtech Summit 11000 probe system. Electrical characterization can be found Figs.~\ref{Figure_4}(a)-(c) and in the Extended Data Figs.~6,~8,~and~10. Electroluminescence measurements were performed by driving the LEDs with a Keithley 4200 parameter analyzer and collecting spectrally resolved light emission from the LED surface (mostly $\mathbf{k} \parallel \mathbf{c}$) by using an Ocean Optics integrating sphere. \\

\section*{Monolithic switching measurements of the HEMT-LED}
Monolithic switching measurements were performed by connecting the source of a HEMT to the anode of a circular LED with a diameter of 400 \textmu m by shorting two electrical probes. A customized probe setup was fabricated to probe both HEMTs and LEDs without the need to flip the sample, as schematized in the Extended Data Fig.~7. The HEMT-LED sample was bonded with the LED facedown to two Ti/Au double-side coated glass slides with thermal glue. The cathode and anode contacts of the LED were subsequently wire-bonded to the
Ti/Au coated slide with a 25 \textmu m diameter Al wire using a West Bond 747630E Wire Bonder. Finally, the coated slides were bonded to two additional uncoated glass slides and the sample was oriented as schematized in the Extended Data Fig.~7 to allow all the contacts of both the HEMT and the LED to be probed from the top. Additionally, to study the backgating effect of the HEMT or to set the backgate voltage, V$_{\mathrm{BS}}$, during switching measurements, indium was soldered to an edge of the sample to make an electrical contact to the bulk n-GaN substrate which sets the backgate potential.
To measure the temporally resolved light output from the LED that was driven by the HEMT, light that was emitted through the surface of the N-polar face, hence passing through the GaN substrate, was measured by a Thorlabs PDA100A Si amplified photodetector. Measurements were performed up to a 5 kHz switching rate which was not a limitation of the dualtronic device but of the photodetector which was set at a gain setting of 70 dB.

\titleformat*{\section}{\Large\bfseries}

\section*{Data availability}

This version of the article has been accepted for publication, after peer review but is not the Version of Record and does not reflect post-acceptance improvements, or any corrections. The Version of Record is available online at: https://doi.org/10.1038/s41586-024-07983-z.

\titleformat*{\section}{\bfseries}

\section*{Acknowledgements}

The work at the Institute of High Pressure Physics PAS was supported by the National Science Centre (NCN) of Poland with Grant No. 2021/43/D/ST3/03266 and the European Union under HORIZON EUROPE Digital, Industry and Space (VISSION ID:101070622).

This work was performed in part at the Cornell NanoScale Facility, a member of the National Nanotechnology Coordinated Infrastructure (NNCI), which is supported by the National Science Foundation (Grant NNCI-2025233). This work further made use of the electron microscopy facility of the Cornell Center for Materials Research (CCMR) with support from the National Science Foundation Materials Research Science and Engineering Centers (MRSEC) program (DMR1719875). The Thermo Fisher Spectra 300 X-CFEG was acquired with support from PARADIM, an NSF MIP (DMR-2039380) and Cornell University. Further funding was provided by the Army Research Office (ARO) with Grant No. W911NF-22-2-0177 and the National Science Foundation with Grant No. DGE-2139899. The authors thank Seohong Park for illustrating Figure~\ref{Figure_3}(e).

\section*{Author contributions}

L.v.D. and E.K. contributed equally. L.v.D. and E.K. were responsible for the process development, device fabrication, device characterization, and monolithic switching measurements. H.T. designed the epistructures and developed the growth process and sequence. M.S. and M.C. were in charge of the HEMT and LED MBE growth. A.F-Z. prepared the substrate and performed atomic force microscopy.  Z.Z. carried out the MBE regrowth of the HEMT source and drain contacts, and performed atomic force microscopy. N.P. and D.M. provided STEM images. D.J., L.v.D., E.K., Z.Z., H.G.X. and H.T. developed the scientific and technological objectives. L.v.D. and E.K. prepared the manuscript. D.J and H.T. revised the manuscript. D.J., H.G.X., and H.T. oversaw the project.

\section*{Competing interests}

The authors filed a US provisional patent (63/548,460) based on the design and technology laid out in this article.

\section*{Additional information}

For correspondence and for questions regarding the materials, L.v.D., E.K., D.J. and H.T. may be contacted.

\end{document}


\graphicspath{{./figures/}}

\title{\large{Extended Data - Leveraging both faces of polar semiconductor wafers for functional devices}}

\author{Len~van~Deurzen*}
\email[Author to whom correspondence should be addressed: ]{lhv9@cornell.edu}
\affiliation{School of Applied and Engineering Physics, Cornell University, Ithaca, New York 14853, USA}
\author{Eungkyun~Kim$^\dagger$}
\affiliation{Department of Electrical and Computer Engineering, Cornell University, Ithaca, New York 14853, USA}
\author{Naomi~Pieczulewski}
\affiliation{Department of Materials Science and Engineering, Cornell University, Ithaca, New York 14853, USA}
\author{Zexuan~Zhang}
\affiliation{Department of Electrical and Computer Engineering, Cornell University, Ithaca, New York 14853, USA}
\author{Anna Feduniewicz-Zmuda}
\author{Mikolaj~Chlipala}
\author{Marcin~Siekacz}
\affiliation{Institute of High Pressure Physics "Unipress", Polish Academy of Sciences, Warsaw 01-142, Poland}
\author{David~Muller}
\affiliation{School of Applied and Engineering Physics, Cornell University, Ithaca, New York 14853, USA}
\author{Huili~Grace~Xing}
\affiliation{Department of Electrical and Computer Engineering, Cornell University, Ithaca, New York 14853, USA}
\affiliation{Department of Materials Science and Engineering, Cornell University, Ithaca, New York 14853, USA}
\affiliation{Kavli Institute at Cornell for Nanoscale Science, Cornell University, Ithaca, New York 14853, USA}
\author{Debdeep~Jena}
\affiliation{School of Applied and Engineering Physics, Cornell University, Ithaca, New York 14853, USA}
\affiliation{Department of Electrical and Computer Engineering, Cornell University, Ithaca, New York 14853, USA}
\affiliation{Department of Materials Science and Engineering, Cornell University, Ithaca, New York 14853, USA}
\affiliation{Kavli Institute at Cornell for Nanoscale Science, Cornell University, Ithaca, New York 14853, USA}
\author{Henryk~Turski}
\affiliation{Department of Electrical and Computer Engineering, Cornell University, Ithaca, New York 14853, USA}
\affiliation{Institute of High Pressure Physics "Unipress", Polish Academy of Sciences, Warsaw 01-142, Poland}

\begin{abstract}
The Extended Data includes:\\
Extended Data Figures 1--10, and References 42-50.
\end{abstract}

\maketitle
\setcounter{page}{12}

\noindent\rule{17.5cm}{0.4pt}

\section*{Extended Data}


\begin{figure}[h]\includegraphics[width=8.9cm]{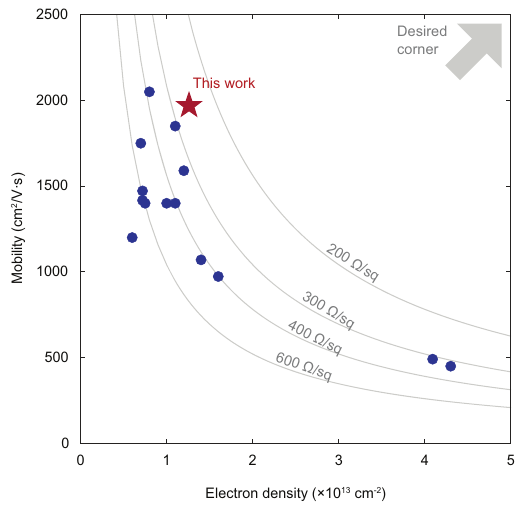}
\caption{Room temperature electron density and mobility of the N-polar HEMTs presented in this study, benchmarked against state-of-the-art N-polar III-nitride HEMTs reported in the literature \cite{diezRecordHighElectron2020,kellerInfluenceSubstrateMisorientation2008,kimFirstDemonstrationNpolar2022,meyerHfO2insulatedGateNpolar2011,pasayatFirstDemonstrationRF2019,rajanNpolarGaNAlGaN2007,singisettiInterfaceRoughnessScattering2012,wongNfaceHighElectron2007,zhangHighdensityPolarizationinduced2D2022}. To the best of our knowledge, the N-polar HEMT investigated in this study demonstrates the lowest reported sheet resistance and one of the highest electron mobilities to date.}
\label{Figure_E1}
\end{figure}

\newpage

\begin{figure}[h]\includegraphics[width=\textwidth]{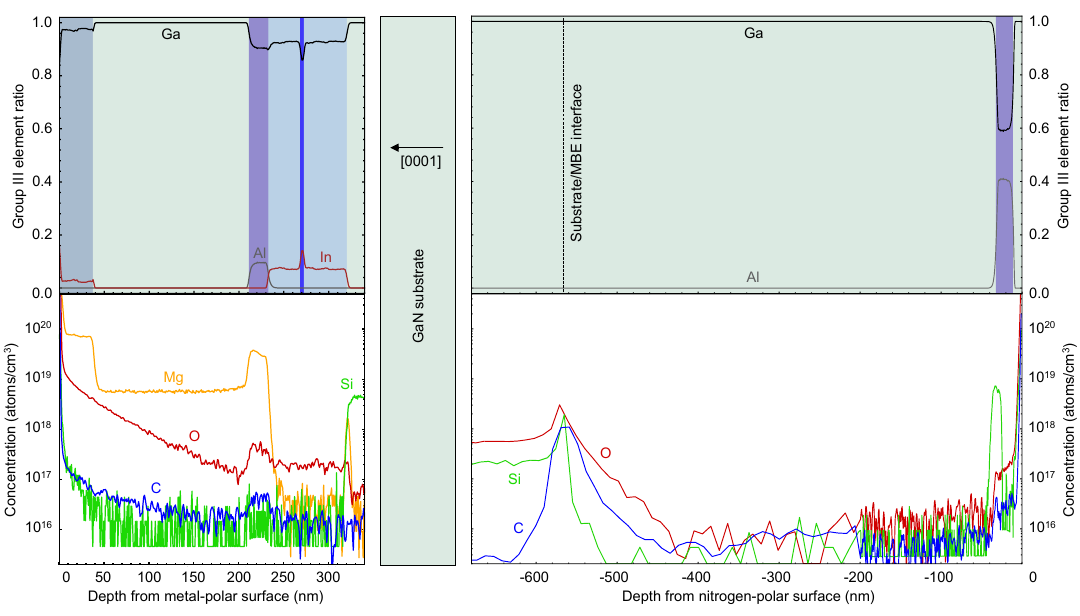}
\caption{Time-of-Flight Secondary-ion Mass Spectrometry (ToF-SIMS). The top row shows the group III element ratios in the LED (left) and HEMT (right) heterostructures, whereas the bottom row shows the Mg, O, Si, and C impurity densities in the same. The detection limits for the impurity densities are $[\mathrm{Mg}] \approx 1-2 \times 10^{16}$~cm$^{-3}$, $[\mathrm{O}] \approx 3-4 \times 10^{16}$~cm$^{-3}$, $[\mathrm{Si}] \approx 7-8 \times 10^{15}$~cm$^{-3}$, and $[\mathrm{C}] \approx 1 \times 10^{16}$~cm$^{-3}$. There is no noticeable interdiffusion of dopants or group-III broadening of the metal-polar LED layers after growth of the N-polar HEMT, indicating interface and doping control is not affected by secondary growth.} 
\label{Figure_E2}
\end{figure}

\newpage

\begin{figure}[h]\includegraphics[width=\textwidth]{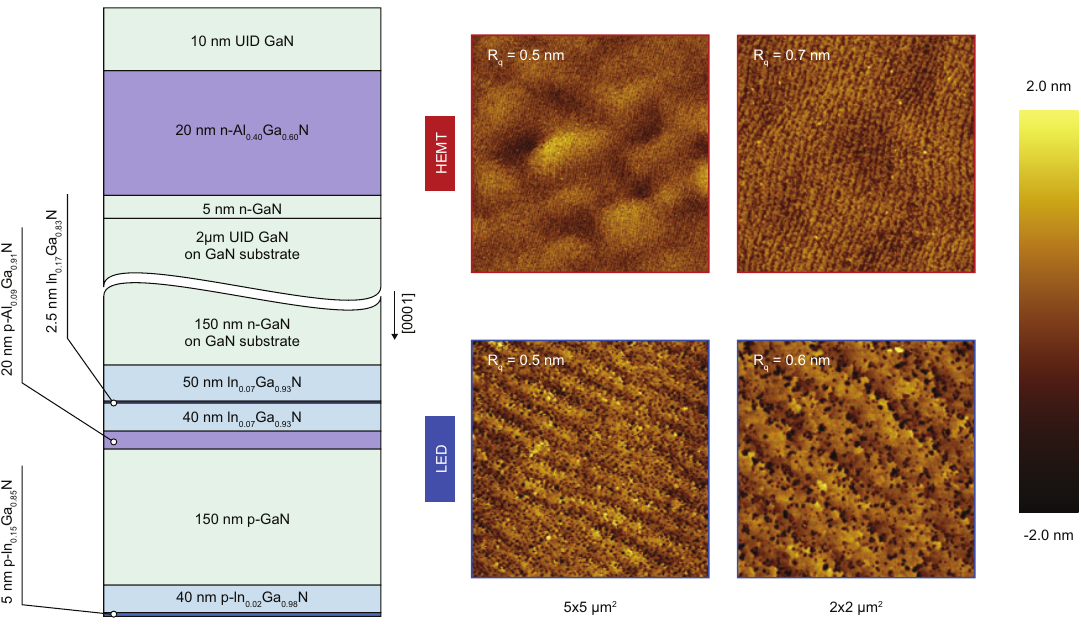}
\caption{$5 \times 5~\upmu$m$^2$ and $2 \times 2~\upmu$m$^2$ atomic force topographs of the $+\mathrm{c}$ and $-\mathrm{c}$ surfaces of the as-grown heterostructures. Sub-nanometer roughness and the observation of trains of well-resolved ondulations are characteristic of step-flow growth for both growth fronts.} 
\label{Figure_E3}
\end{figure}

\newpage

\begin{figure}[h]\includegraphics[width=8.9cm]{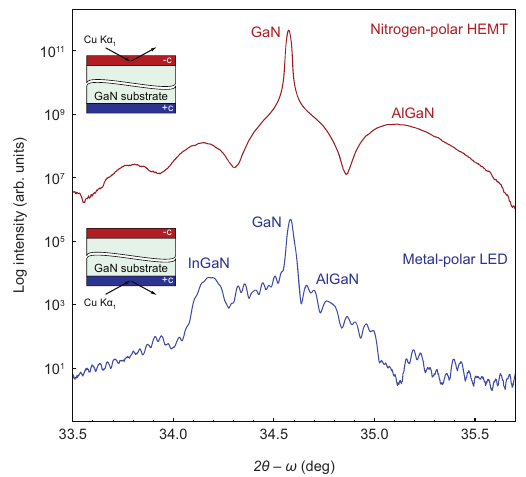}
\caption{Semilog plot of 2$\theta - \omega$ XRD scans around the 002 reflections corresponding to the nitrogen-polar HEMT (top, red) and metal-polar LED (bottom, blue). Each y-axis tick represents two orders in detector counts. Besides the sharp and intense peak at 34.57$^{\circ}$ due to the GaN substrate, higher-angle AlGaN related peaks are present for both the HEMT and LED, whereas InGaN related peaks are present at lower angles for the LED. The locations and FWHMs of the peaks along with the observation of interference fringes indicate that the heterostructure interfaces are atomically sharp and that the layers are coherently strained to the GaN substrate.}
\label{Figure_E4}
\end{figure}

\newpage

\begin{figure}[h]\includegraphics[width=\textwidth]{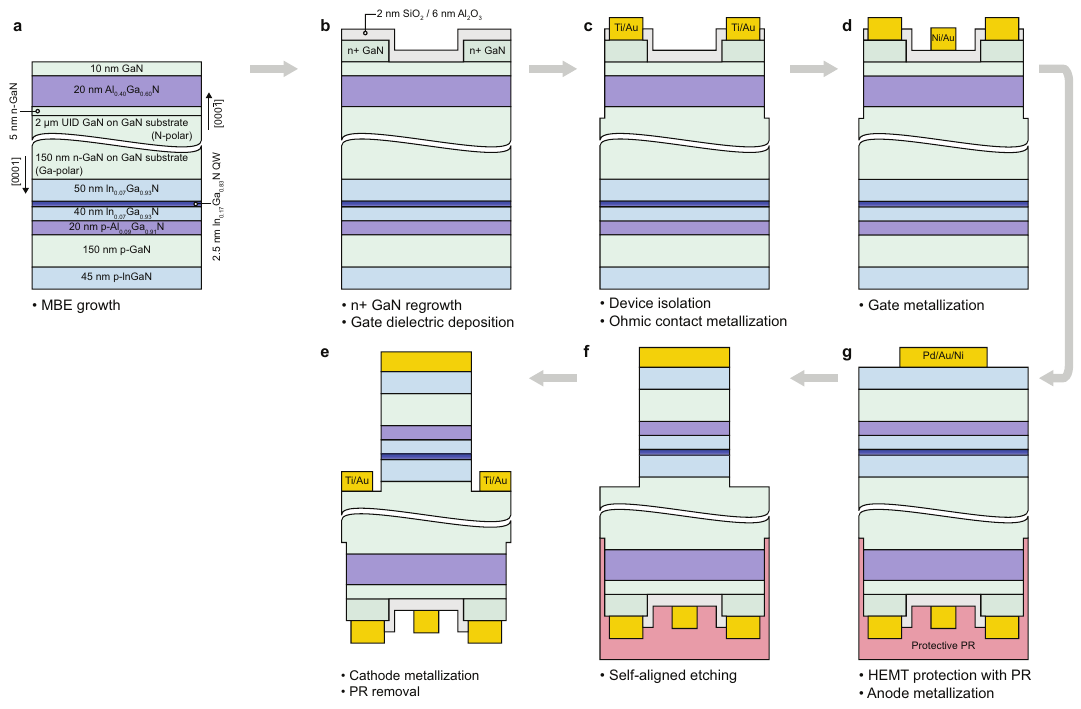}
\caption{Detailed process flow of the HEMT-LED. Starting from the as-grown heterostructures, the gray arrows follow the processing steps chronologically, with the metal-polar LED being fabricated after the N-polar HEMT. The process details are described in the methods section.}
\label{Figure_E5}
\end{figure}

\newpage
\begin{figure}[h]\includegraphics[width=\textwidth]{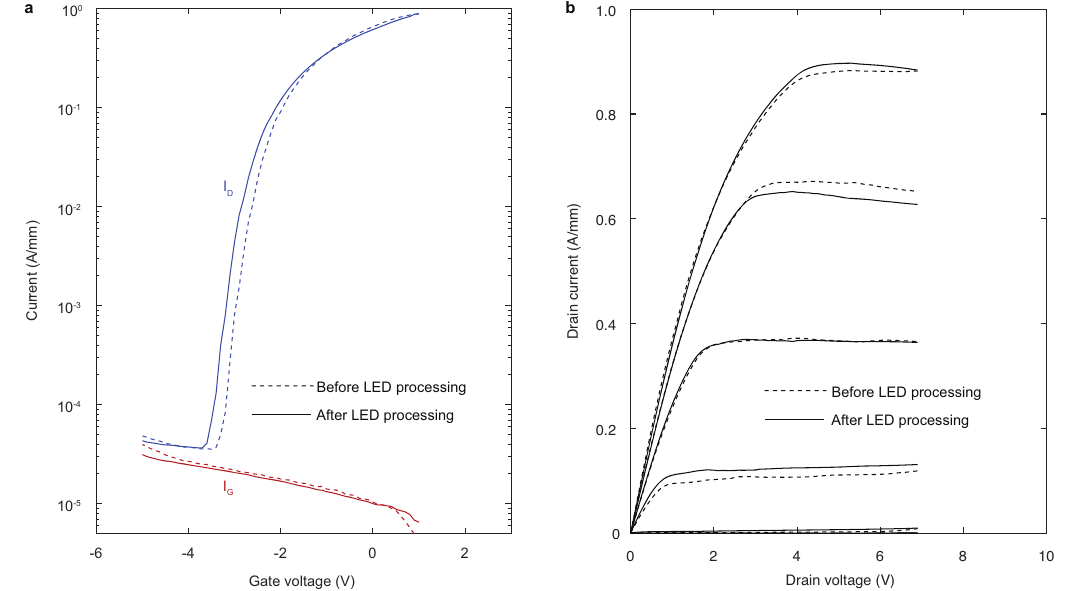}
\caption{Feasibility of dualtronic device processing. (a) Log scale transfer curves and (b) output curves of a N-polar HEMT before (dashed) and after (solid) the LED fabrication, showing a negligible change in gate leakage current, on-resistance, and output current. The shift in threshold voltage after the LED fabrication is estimated to be $\sim$~0.2~V.} 
\label{Figure_E6}
\end{figure}

\newpage

\begin{figure*}[h]\includegraphics[width=\textwidth]{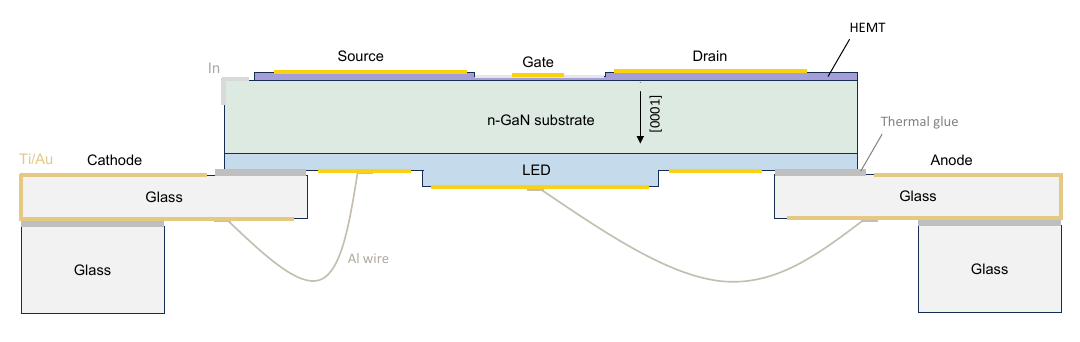}
\caption{Schematic of top-side probing setup used for HEMT backgating and HEMT-LED switching measurements. The HEMT-LED sample was bonded with the LED facedown to two Ti/Au double-side coated glass slides with thermal glue. The cathode and anode contacts of a 400 \textmu m diameter LED were subsequently wire-bonded to the Ti/Au coated slide with a 25~\textmu m diameter Al wire. Finally, the coated slides were bonded to two additional uncoated glass slides and and the sample was oriented as schematized to allow all the contacts of both the HEMT and the LED to be probed from the top. Additionally, to study the backgating effect of the HEMT or to set the body voltage during switching measurements, indium was soldered to an edge of the sample to make an electrical contact to the bulk n-GaN substrate. To facilitate the access of both device fronts by establishing a reliable electrical connectivity between both substrate faces, the use of through-hole contacts is currently being explored.} 
\label{Figure_E7}
\end{figure*}

\newpage

\begin{figure}[h]\includegraphics[width=\textwidth]{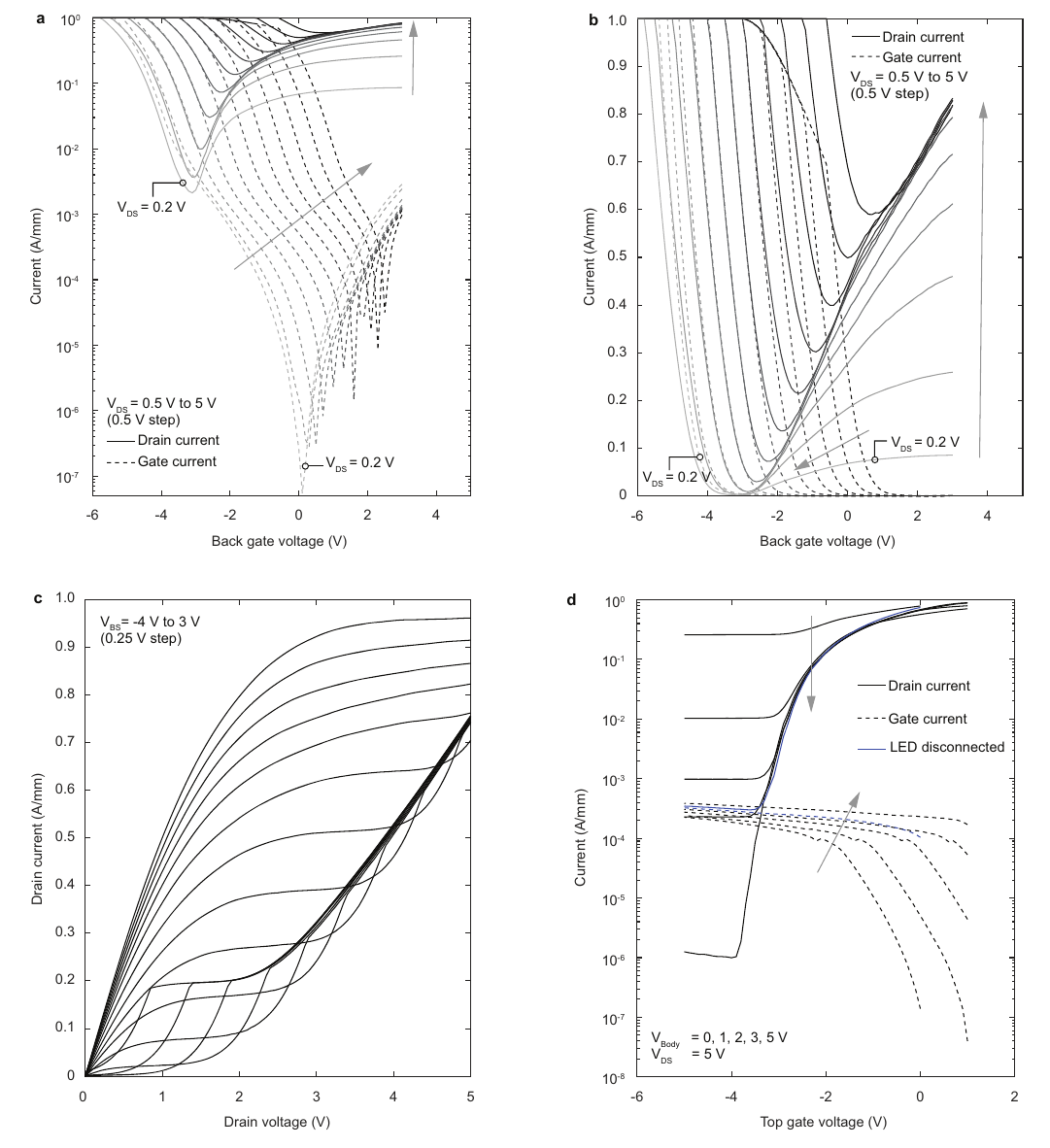}
\caption{Backgating effect of N-polar HEMT. (a) Log scale and (b) linear scale transfer curves of a three-terminal N-polar HEMT gated by the substrate contact made with Indium as shown in Extended Data Figure~1, with the top gate left floating. Each curve represents the dependency of the drain (solid) and gate (dashed) current as a function of the back gate voltage (V\textsubscript{BS}), with a fixed drain bias. Drain bias increases from 0.5~V to 5~V at a 0.5~V step in the direction the arrow points, with the last curve taken at a drain bias (V\textsubscript{DS}) of 0.2~V. (c) Output characteristics of a back-gated N-polar HEMT. The maximum drain current is comparable to the value measured on the top-gated HEMT. For V\textsubscript{BS} - V\textsubscript{DS} $\lesssim$ -4~V, the drain current is dominated by the high gate leakage current. (d) Log scale transfer curves of a four-terminal N-polar HEMT measurement with fixed V\textsubscript{BS} and V\textsubscript{DS}. The blue curve represents the transfer curve with a back gate. No drain current modulation is observed for V\textsubscript{DS} - V\textsubscript{BS} $\sim$ 5~V, and up to six orders of drain current modulation is observed for V\textsubscript{DS} - V\textsubscript{BS} = 0~V.}
\label{Figure_E8}
\end{figure}

\newpage

\begin{figure}[h]\includegraphics[width=8.9cm]{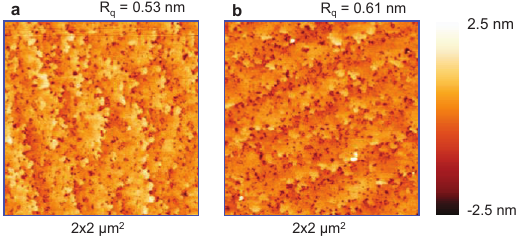}
\caption{2~$\times$~2 \textmu m$^{2}$ atomic force topography of the +$c$ (LED) surface after (a) LED growth and (b) LED and HEMT growth. The +$c$ surface morphology of the LED is not affected by the secondary N-polar growth procedures, including gallium mounting, HEMT growth, and wet chemical piranha treatment.}
\label{Figure_E9}
\end{figure}

\clearpage
\newpage 

\begin{figure}[h]\includegraphics[width=8.9cm]{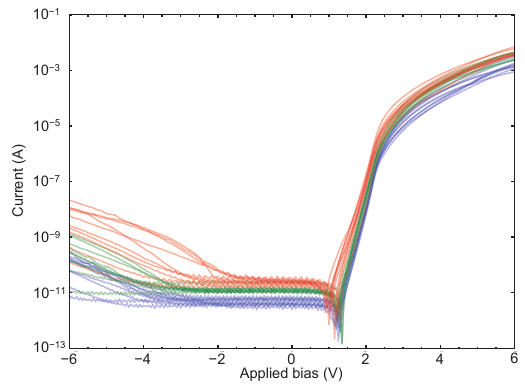}
\caption{IV characteristics of ten circular diodes with 50~\textmu m diameter (blue), ten circular diodes with 100~\textmu m diameter (green), and ten circular diodes with 140~\textmu m diameter (red), measured at an applied bias ranging from -6~V to 6~V. At ±6 V, all measured diodes show rectifying behavior in the range of $2.8 \times 10^5$ to $2.9 \times 10^8$, with an average of $2.8 \times 10^7$. Current modulation over 11 orders is achieved in the diodes for the applied voltage range, with the minimum current limited by the measurement setup. The electrical turn-on of the diodes is around 1.4 V for all measured devices, and all LEDs exhibit light emission characteristics with spectra comparable to Figure~4(d) in the main manuscript with detectable emission above 2.5-3~V forward bias.}
\label{Figure_E10}
\end{figure}

\clearpage

\bibliography{supplement}